\documentclass[]{aa}
\usepackage{siunitx}
\usepackage{graphicx}   
\usepackage[hyperfootnotes=true, hidelinks, colorlinks, citecolor=blue, linkcolor=blue, linktocpage, bookmarks=true]{hyperref}
\usepackage{txfonts}
\usepackage{soul}
\usepackage{multicol, blindtext}

\newcommand{\zem}{$z_{\rm em}$}

\newcommand{\kms}{km\,s$^{-1}$}
\newcommand{\cmsq}{${\rm cm}^{-2}$}

\newcommand{\hi}{\mbox{H\,{\sc i}}}
\newcommand{\civ}{\mbox{C\,{\sc iv}}}
\newcommand{\mgii}{\mbox{Mg\,{\sc ii}}}

\begin{document} 

   \title{MALS discovery of a rare \hi\ 21-cm absorber at $z\sim1.35$:\\ origin of the absorbing gas in powerful AGN}
   \titlerunning{Cold gas associated with J2339-5523}
   \authorrunning{Deka, P. P. et al.}

   \author{P. P. Deka \href{https://orcid.org/0000-0001-9174-1186}{\includegraphics[width=8pt]{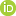}}
          \inst{1}
          \and
          N.~Gupta \href{https://orcid.org/0000-0001-7547-4241}{\includegraphics[width=8pt]{orcid.png}}\inst{1} \and H.~W.~Chen\inst{2} \and S.~D.~Johnson\inst{3} \and P.~Noterdaeme\inst{4,5} \and F.~Combes\inst{6} \and E.~Boettcher \inst{7,8,9} \and S.~A.~Balashev\inst{10,11} \and K.~L.~Emig \href{https://orcid.org/0000-0001-6527-6954}{\includegraphics[width=8pt]{orcid.png}}\fnmsep\thanks{Jansky Fellow of the National Radio Astronomy Observatory} \inst{12} \and G.~I.~G.~J\'ozsa\inst{13,14} \and H.-R.~Kl\"ockner\inst{13} \and J-.K.~Krogager\inst{15} \and E.~Momjian \href{https://orcid.org/0000-0003-3168-5922}{\includegraphics[width=8pt]{orcid.png}}\inst{16} \and P.~Petitjean\inst{4} \and G.~C.~Rudie \inst{17} \and J.~Wagenveld \inst{13} \and F.~S.~Zahedy \inst{17}
          }

   \institute{Inter-University Centre for Astronomy and Astrophysics, Post Bag 4, Ganeshkhind, Pune 411 007, India\\
              \email{\href{parthad@iucaa.in}{parthad@iucaa.in}}
         \and
             Department of Astronomy \& Astrophysics, The University of Chicago, 5640 South Ellis Avenue, Chicago, IL 60637, USA
        \and
            Department of Astronomy, University of Michigan, Ann Arbor, MI 48109, USA
        \and
            Institut d'Astrophysique de Paris, UMR 7095, CNRS-SU, 98bis bd Arago, 75014  Paris, France
        \and
             Franco-Chilean Laboratory for Astronomy, IRL 3386, CNRS and Departamento de Astronom\'ia, Universidad de Chile, Casilla 36-D, Santiago, Chile
        \and
            Observatoire de Paris, Coll\`ege de France, PSL University, Sorbonne University, CNRS, LERMA, Paris, France
        \and
            Department of Astronomy, University of Maryland, College Park, MD 20742, USA
        \and
            X-ray Astrophysics Laboratory, NASA/GSFC, Greenbelt, MD 20771, USA
        \and   
            Center for Research and Exploration in Space Science and Technology, NASA/GSFC, Greenbelt, MD 20771, USA
        \and
            Ioffe Institute, 26 Politeknicheskaya st., St. Petersburg, 194021, Russia
        \and
            HSE University, Saint Petersburg, Russia
        \and
            National Radio Astronomy Observatory, 520 Edgemont Road, Charlottesville, VA 22903, USA
        \and
            Max-Planck-Institut f\"ur Radioastronomie, Auf dem H\"ugel 69, D-53121 Bonn, Germany
        \and
            Department of Physics and Electronics, Rhodes University, P.O. Box 94 Makhanda 6140, South Africa
        \and
            Universit\'e Lyon1, ENS de Lyon, CNRS, Centre de Recherche Astrophysique de Lyon UMR5574, F-69230 Saint-Genis-Laval, France
        \and
            National Radio Astronomy Observatory, 1003 Lopezville Rd., Socorro, NM 87801, USA
        \and
            The Observatories of the Carnegie Institution for Science, 813 Santa Barbara Street, Pasadena, CA 91101, USA
        }

    \date{Received \today; accepted bbb}
    \mail{hliszt@nrao.edu}%

 
  \abstract
   {We report a new, rare detection of \hi\ 21-cm absorption associated with a quasar (only six known at $1<z<2$) here towards J2339-5523 at \zem = 1.3531, discovered through the MeerKAT Absorption Line Survey (MALS). The absorption profile is broad ($\sim 400$~\kms\,), and the peak is redshifted by $\sim 200$~\kms\, from \zem\,. Interestingly, optical/FUV spectra of the quasar from Magellan-MIKE/HST-COS spectrographs do not show any absorption features associated with the 21-cm absorption. This is despite the coincident presence of the optical quasar and the radio `core' inferred from a flat spectrum component of flux density $\sim 65$ mJy at high frequencies ($> 5$ GHz).  The simplest explanation would be that no large \hi\ column ($N$(\hi) $>10^{17}$\,\cmsq) is present towards the radio ‘core’ and the optical AGN. Based on the joint optical and radio analysis of a heterogeneous sample of 16 quasars ($z_{median}$ = 0.7) and 19 radio galaxies ($z_{median}$ =  0.4) with \hi\ 21-cm absorption detection and matched in 1.4\,GHz luminosity (L$_{\rm 1.4\,GHz}$), a consistent picture emerges where quasars are primarily tracing the gas in the inner circumnuclear disk and cocoon created by the jet-ISM interaction. These exhibit L$_{1.4\,\rm GHz}$ -- $\Delta V_{\rm null}$ correlation, and frequent mismatch between the radio and optical spectral lines.  The radio galaxies show no such correlation and likely trace the gas from the cocoon and the galaxy-wide ISM outside the photoionization cone. The analysis presented here demonstrates the potential of radio spectroscopic observations to reveal the origin of the absorbing gas associated with AGN that may be missed in optical observations.}


\maketitle
%
\section{Introduction} 
\label{sec:intro}  

Both observations and simulations provide strong support for the feedback from active galactic nuclei (AGN) in regulating the star formation efficiency of the host galaxy \citep[e.g.,][]{Croton06, Fabian12}.   In this context, of particular importance is the impact of AGN feedback on the physical state of cold atomic and molecular gas, which represent the raw material for star formation and directly affect the evolutionary track of the galaxy \citep[e.g.,][]{Veilleux20}.
Absorption line studies provide  a powerful way to study these phases in the AGN environment. At optical wavelengths, obscuration of the AGN by dust prevents detection of dense molecular clouds in absorption. In fact, mostly diffuse (presumably warm, $\sim 10^4$~K) \hi\ has been detected through proximate damped Lyman-$\alpha$ absorption \citep[e.g.][]{Ellison2010} but progress is being done to reach colder ($\sim$100\,K) and denser phases using H$_2$  observations \citep{Noterdaeme2019}. At radio wavelengths, \hi\ 21-cm absorption lines in turn provide a dust-unbiased way to probe the cold atomic gas towards the population of radio-loud AGNs.

To date, more than 500 AGN have been searched for \hi\ 21-cm absorption at $0<z<5$ \citep[see][for a review]{Morganti18}.
The majority of \hi\ 21-cm absorption detections are associated with Compact Steep Spectrum (CSS) and Gigahertz Peaked Spectrum (GPS) radio sources (sizes $<$15\,kpc) at low redshifts ($z<0.3$). The typical \hi\ column densities for an assumed spin temperature of 100\,K in the case of detections are $10^{20 - 21}$\,\cmsq.   The high 21-cm absorption detection rate ($\sim$30\%) among these is believed to be a direct consequence of the fact that these radio sources are young (age $<10^5$\,yrs) and still embedded in the interstellar medium (ISM) of gas-rich host galaxy \citep[e.g.,][]{Vermeulen03, Gupta06}. This scenario is also supported by the observational signatures of gas outflows occurring at sub-kpc scales resulting from radio jet-ISM interactions detected through an excess of blue-shifted absorption lines \citep[e.g.,][]{Morganti2005b,Morganti05, Mahony13, Gereb15, Morganti18}.
Further, a bulk of 21-cm absorption detections are associated with type\,{\sc II} AGNs (i.e., here radio-galaxies) that are seen edge-on, leading to the conclusions that a significant portion of the absorbing gas, especially those represented by strong narrow absorption components, has origins in the obscuring torus or galaxy-wide ISM outside the photoionization cone \citep[][]{Pihlstrom03, Gupta06, Gupta06uni}.


Quantifying evolution with redshift in the properties of cold gas associated with AGN ought to be an obvious goal of \hi\ 21-cm absorption line studies.  However, despite major efforts over the last three decades, only a handful of absorption systems are known at $z>1$: six at $1<z<2$ and four at $2<z<3.5$ \citep[e.g.,][]{Aditya18, Gupta21hz, Curran12uv}. 
Specifically, the majority of targets at $z>2$ are powerful quasars ($\log \rm L_{1.4\rm GHz} \rm (W\,Hz^{-1}) \gtrsim 27.0$).  These show an extremely small \hi\ 21-cm absorption detection rate ($1.6^{+3.8}_{-1.4}$\%), which is not at odds with the observed rate (2-5\%) of proximate ($\Delta v < 3000$ \kms\ from $z_{em}$) damped Ly$\alpha$ absorbers \citep[$N$(\hi)$>10^{20.3}$\,\cmsq;][]{Ellison02}.
Overall, no firm conclusions can be drawn about the redshift evolution, due to the small number of high-$z$ detections,  different selection methods applied to low- and high-$z$ samples and luminosity bias affecting uniformly selected samples \citep[see][]{Aditya18gps, Aditya18, Grasha19}. 

%

Here, we report the first associated \hi\ 21-cm absorption detection discovered by the ongoing MeerKAT Absorption Line Survey \citep[MALS;][]{Gupta17mals}. The corresponding AGN at $z=1.3531$ is a bonafide quasar \citep[][]{Veron-cetty2001} and an extremely powerful radio source with 1.4\,GHz luminosity, L$_{\rm 1.4GHz}$ = $10^{27.2}$\,W\,Hz$^{-1}$.  The combination of compact radio emission \citep[$<$5.2\,kpc;][]{Boettcher22} and the steep spectral index ($\alpha\sim$$-$0.8) - modeled using a power law, $S_\nu \propto \nu^\alpha$ -  imply that the radio AGN may be classified as a CSS source \citep[][]{Odea21}. Interestingly, no signatures of neutral gas at the \hi\ 21-cm absorption redshift are detected in the optical-ultraviolet spectrum.  The joint radio-optical analysis of this case allows us to discuss the origin and distribution of absorbing gas in the vicinity of the radio source.

 This paper is structured as follows. In Section~\ref{sec:obs}, we present details of the radio and FUV / optical observations and their data analysis. In Section~\ref{sec:res}, we present the results and explore the possibility of the absorption occurring against the radio-core component. In Section~\ref{sec:QSOvsGAL}, we discuss J2339$-$5523 in the context of other quasars and radio galaxies in the literature with reported \hi\ 21-cm detections and draw general conclusion regarding the origin of the absorbing gas. The results are summarized in Section~\ref{sec:summary}.

Throughout this paper we use a flat $\Lambda$ cosmology with $\Omega_m$=0.315, $\Omega_\Lambda$=0.685 and H$_{\rm 0}$=67.4\,\kms\,Mpc$^{-1}$ \citep[][]{Planck20}. At $z = 1.3531$, $1^{\prime\prime}$ = 8.642\,kpc.

\section{Observations and data analysis}      
\label{sec:obs} 

\subsection{Radio Observations}
\label{sec:mrktobs} 

J2339-5523 was observed with the MeerKAT-64 array, using the 32K mode of the SKA Reconfigurable Application Board (SKARAB) correlator, on 2020 June 14 and 22 (L-band; 856 - 1712\,MHz), and 2021 January 07 (UHF-band; 544 - 1088\,MHz).  The details of the L-band datasets providing a total on-source time of 112\,mins are presented in \citet[][]{Boettcher22}. 
For the UHF observations, the SKARAB correlator was configured to split the 544\,MHz bandwidth centered at 815.9917\,MHz into 32768 frequency channels. We also observed J0408-6545 for flux density scale, delay and bandpass calibrations, and J2329-4730 for complex gain calibration.  The data were acquired in all 4 polarization products XX, XY, YX and YY.  The correlator dump time was 8 seconds. The total integration time on J2339-5523 was 120\,mins.

\begin{figure} 
\centerline{\vbox{
\centerline{\hbox{ 
\includegraphics[
trim = {0cm 0cm 0cm 0cm}, clip=true,
width=0.45\textwidth,angle=0]{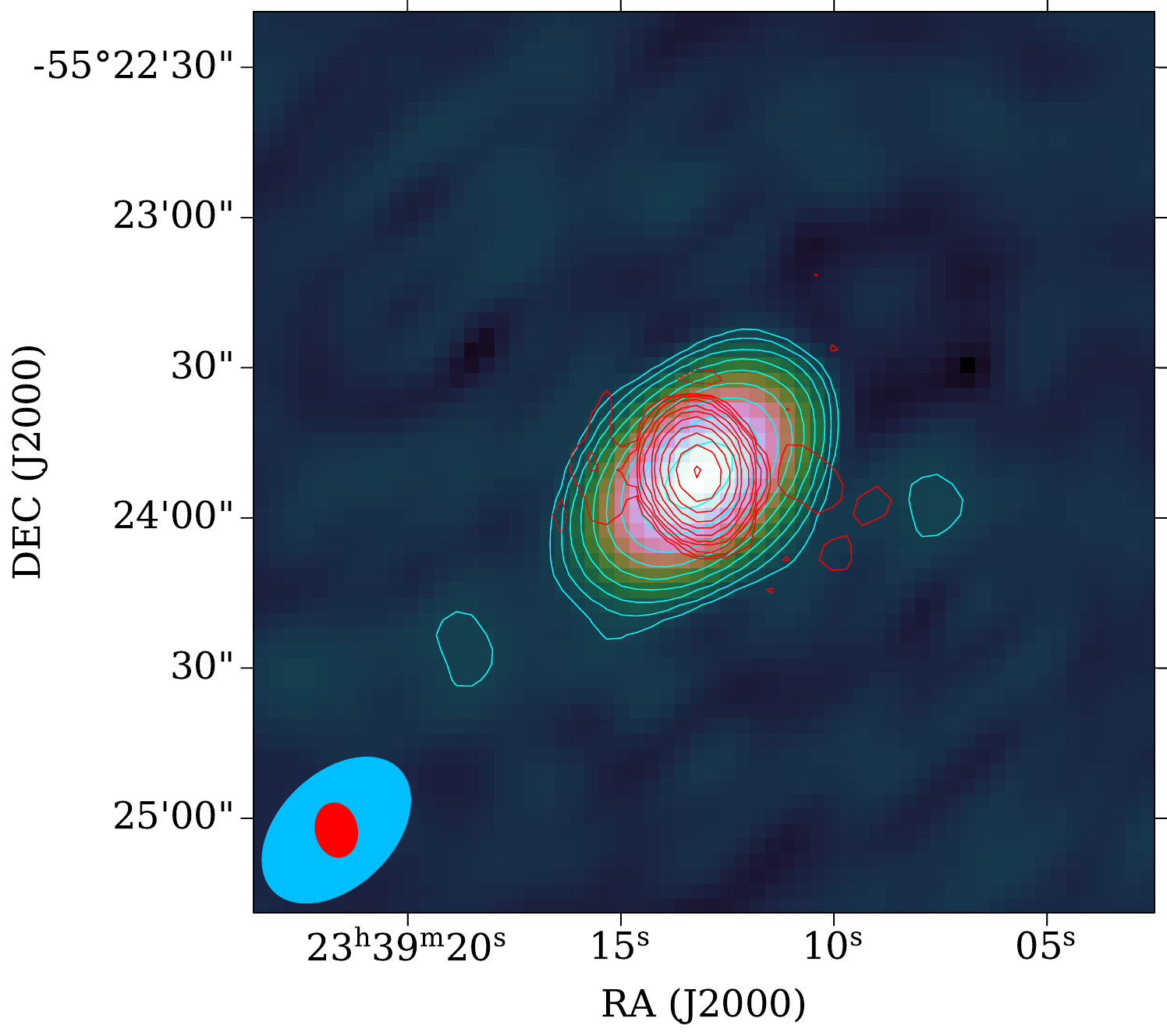}
}} 
}}  
\vskip+0.0cm  
\caption{
A portion of the larger (6k\,$\times$\,6k pixels) MeerKAT Stokes-$I$ continuum images centered on the quasar J2339-5523.  The background is UHF-band image at 580.1\,MHz (SPW0).  The contour levels are n $\times$ (-1, 1, 2, 4, 8, 16, ...), where n = 1.0\,mJy\,beam$^{-1}$ (cyan) and 0.15\,mJy\,beam$^{-1}$ (red) for the UHF- and L-band images at 580.1\,MHz (SPW0) and 1642.3\,MHz (SPW14), respectively. The restoring beams are shown as ellipses at the bottom left corner of the image. The shapes of cyan and red contours simply follow the corresponding restoring beams, as is expected for the unresolved emission (see text for details).}

\label{fig:contim}   
\end{figure}


\begin{figure} 
\centerline{\vbox{
\centerline{\hbox{ 
\includegraphics[
trim = {0cm 0cm 0cm 0cm}, clip=true,
width=0.50\textwidth,angle=0]{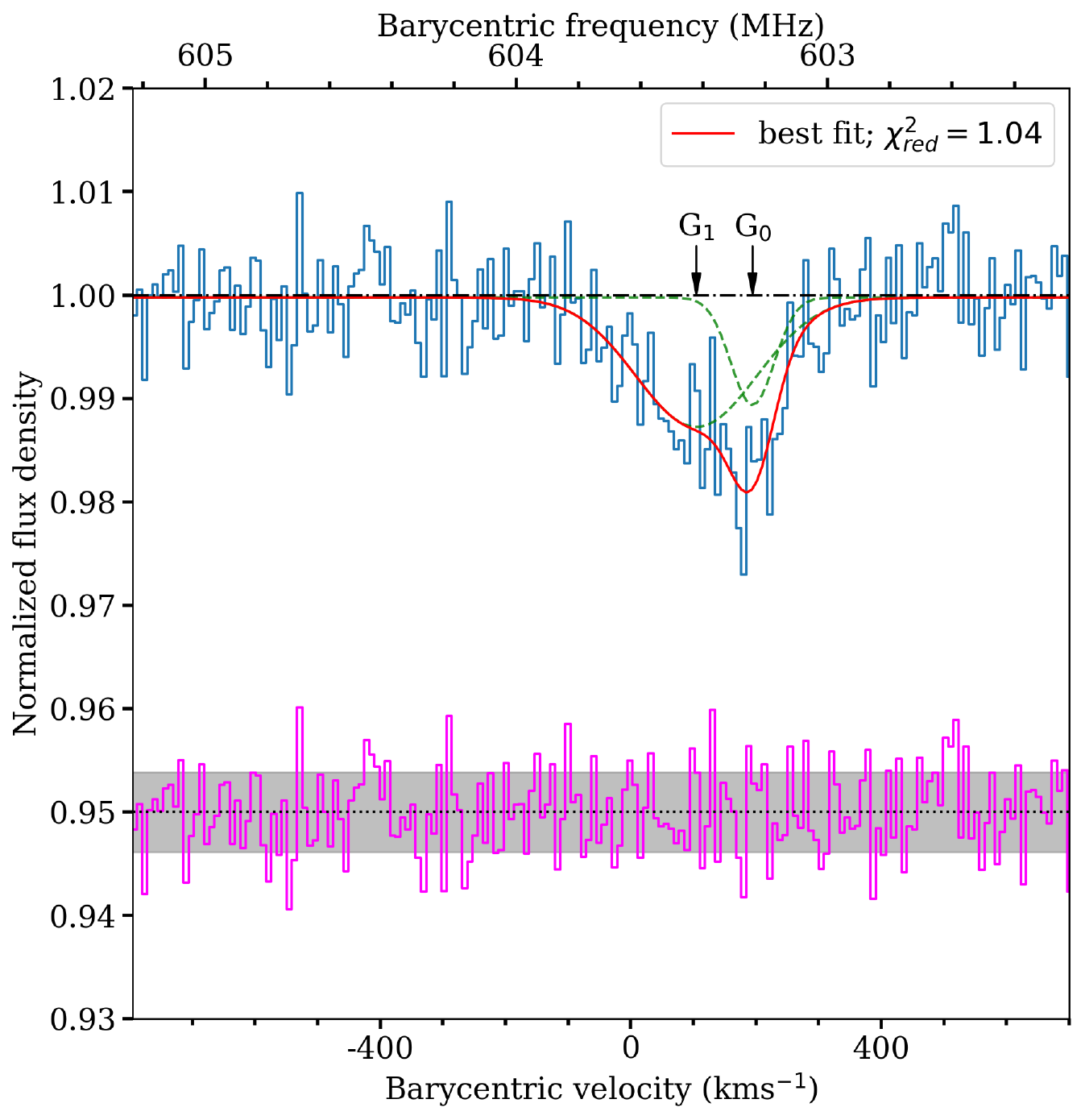}  
}} 
}}  
\vskip+0.0cm  
\caption{
\hi\ 21-cm absorption associated with J2339-5523. Zero velocity corresponds to \zem = 1.3531.   Individual Gaussian components (Table~\ref{tab:21cmfit}) and the combined best-fit model are overplotted as dashed (green) and continuous lines (red), respectively.  The residuals from the fit (magenta curve) and the $1\sigma$ uncertainty on normalized flux density (shaded region in grey), arbitrarily offset for clarity, are also shown. 
}
\label{fig:21cm}   
\end{figure}

\begin{table}
\caption{Multiple Gaussian fits to the \hi\ absorption in J2339-5523. }
\vspace{-0.4cm}
\begin{center}
\begin{tabular}{ccccc}
\hline
\hline
{\large \strut}        &    $z_{abs}$  &    Centre     &  $\sigma$            &   $\tau_p$       \\
             Id.\      &                &   (\kms)      &   (\kms)             &   ($10^{-3}$)    \\
\hline
        \multicolumn{5}{c}{Stokes I; $\chi^2_{red} \sim 1.04$} \\
\hline
                G$_0$ & 1.3546 $\pm$ 0.0002    &  195 $\pm$ 35       & 34 $\pm$ 11       &    10.4 $\pm$ 3.2       \\
                G$_1$ & 1.3539 $\pm$ 0.0004     &    105 $\pm$ 23       &  98 $\pm$ 13       &    12.8 $\pm$ 1.9        \\
\hline

\end{tabular}
\label{tab:21cmfit}
\end{center}
\end{table}

The data were flagged, calibrated and imaged using the Automated Radio Telescope Imaging Pipeline (ARTIP) described in \citet[][]{Gupta21}. The details of  Stokes-$I$ continuum and spectral line imaging steps specific to L- and UHF-band datasets are presented in \citet[][]{Gupta21} and \citet[][]{Combes21}, respectively. 
The spectral line processing in ARTIP proceeds by splitting the band into 15 spectral windows, labeled as SPW0 to SPW14, and generates continuum images and spectral line image cubes for each of these.   
For both continuum and spectral line imaging we have used Briggs weighting with a robust factor of 0 as implemented in the Common Astronomy Software Applications \citep[CASA;][]{Mcmullin07}. 
In Fig.~\ref{fig:contim}, we show Stokes-$I$ continuum images of the lowest and highest frequency spectral windows from UHF and L bands, respectively. The restoring beams for these images at 580.1\,MHz and 1642.3\,MHz are $23\farcs2\times15\farcs0$ (position angle = $-46.2^\circ$)  and $7\farcs2\times5\farcs5$ (position angle = $+11.8^\circ$), respectively. The corresponding flux densities at these frequencies i.e., 363.7$\pm$0.6\,mJy at 580.1\,MHz and 166.0$\pm$0.1\,mJy at 1642.3\,MHz lead to a spectral index of $\alpha_{580}^{1642}$ = -0.753 $\pm$ 0.002. The rms noise values close to the central bright source are 300\,$\mu$Jy\,beam$^{-1}$ and 50\,$\mu$Jy\,beam$^{-1}$, respectively. 
The radio emission is unresolved in these images, and using the 1642.3\,MHz image we estimate a deconvolved size $<0\farcs6$ \citep[see also][]{Boettcher22}. 
At $z = 1.35$, the 21-cm line frequency falls in the SPW1 of the UHF-band, covering 594.983 -- 633.233\,MHz and centered at 614.1\,MHz (restoring beam = $21\farcs6\times14\farcs3$; position angle = $-45.7^\circ$). Using a single component Gaussian fit to the radio emission, we estimate the total continuum and peak flux density of the source to be 343.0 $\pm$ 0.7\,mJy and 336.4 $\pm$ 0.5\,mJy\,beam$^{-1}$, respectively. 

Since the radio emission associated with J2339-5523 is unresolved in the MeerKAT images, we extracted spectra at the pixel ($3^{\prime\prime}$) corresponding to the peak intensity. 
As shown in Fig.~\ref{fig:21cm}, we clearly detect \hi\ 21-cm absorption associated with J2339-5523. At the detected frequency, based on the spectra of other radio sources observed in MALS there is no known RFI, and the absorption is consistently detected in both XX and YY spectra. The absorption profile can be well modeled (reduced $\chi^2 = 1.04$) with two Gaussians. The details of the fit and the values of the best-fit parameters are given in Table~\ref{tab:21cmfit}. The velocity width between the nulls of the absorption profile ($\Delta V_{\rm null}$) is $\approx$ 400 \kms.
The channel width and rms noise in the vicinity of the redshifted \hi\ 21-cm line frequency in the unsmoothed spectrum are 8.3\,\kms\ and 1.1\,mJy\,beam$^{-1}$\,channel$^{-1}$, respectively.  
The spectrum presented in Fig.~\ref{fig:21cm} has been normalized using the peak flux density of the continuum emission.

\subsection{Far-ultraviolet and optical absorption spectra}  
\label{sec:mikeobs} 

Medium-resolution, high signal-to-noise ratio ($S/N$) far-ultraviolet (FUV) spectra of J2339$-$5523 were obtained under the Cosmic Ultraviolet Baryon Survey (CUBS) program (PID$=$15163; PI: Chen) using the Cosmic Origins Spectrograph (COS; \citealt{Green2012}) on board the Hubble Space Telescope (HST).  The COS FUV spectra offer a contiguous spectral coverage of $\lambda = 1100$–1800 \AA, with full width at half-maximum (${\rm FWHM}$) of the instrumental function of $\approx 20$ \kms.  Optical Echelle spectra of J2339$-$5523 was obtained using the Magellan Inamori Kyocera Echelle \citep[MIKE;][]{Bernstein2003} spectrograph on the Magellan Clay Telescope, which cover a wavelength range of $\lambda = 3330$–9300~\AA\ with an ${\rm FWHM}\approx 8$~\kms.  Details regarding the observations and data reduction of MIKE spectra are presented in \cite{Chen:2020aa}.  All spectra are continuum-normalized using a low-order polynomial fit to spectral regions free of strong absorption features.

The final combined COS spectrum of the quasar has a median signal-to-noise ($S/N$) per resolution element of $\langle\,S/N\rangle_{\rm resol}=22$ over the FUV spectral window.  The final combined optical echelle spectrum has a $\langle\,S/N\rangle_{\rm resol}=15$ at $\lambda\approx 3500$ \AA. The joint COS FUV and MIKE optical echelle spectra of this quasar provide extended spectral coverage for constraining the abundances of neutral gas and associated metal ions along the line of sight.  Specifically at $z=1.3531$, sensitive constraints can be obtained for \ion{He}{i}, \ion{C}{ii}, \ion{C}{iv}, \ion{O}{iii}, \ion{Al}{ii}, \ion{Al}{iii}, \ion{Mg}{ii}, \ion{Fe}{ii}, \ion{Ca}{ii}.  However, the hydrogen Lyman series lines are redshifted outside of the COS FUV window.  No direct UV absorption constraints are available for \hi\,.

\section{Results}    
\label{sec:res}  

\subsection{\hi\ 21-cm absorption line gas}
\label{sec:hiabs}

The neutral hydrogen column density $N$(\hi), spin temperature $T_{\rm s}$, and covering factor of the radio emission by the absorbing gas ($f_c$) are related to the integrated optical depth ($\int\tau dv$), through
\begin{equation}
	N{(\hi)}=1.823\times10^{18}~{T_{\rm s}\over f_{\rm c}}\int~\tau(\rm v)~{\rm dv}~{\rm cm^{-2}}.
\label{eq21cm}
\end{equation}
The absorption profile presented in Fig.~\ref{fig:21cm} corresponds to $\int\tau$dv = 3.7$\pm$0.2\,\kms.  Assuming $T_{\rm s}$=100\,K and $f_c$=1, the total column density is (6.7$\pm$0.4)$\times10^{20}(T_S / 100)(1/f_c)$\,\cmsq.  
The spin temperature is directly controlled by collisions, absorption of 21-cm continuum radiation and pumping by Ly$\alpha$ photons \citep[][]{Field58}.  In the neighborhood of AGN, the radiative effects may dominate and $T_{\rm s}\gg100\,K$ \citep[][]{Maloney96}.  In order to derive $N$(\hi), besides $T_{\rm s}$, we also need constraints on the fraction of radio continuum emission covered by the absorbing gas i.e., $f_c$ which may be less than 1.  The assumptions of $T_{\rm s}$ = 100\,K and $f_c$ = 1 therefore provide a lower limit to the true $N$(\hi).

\subsection{Comparison with optical/UV spectral line}
\label{sec:optlines}

\begin{figure} 
\centerline{\vbox{
\centerline{\hbox{ 
\includegraphics[
trim = {0cm 0cm 0cm 0cm}, clip=true,
width=0.50\textwidth,angle=0]{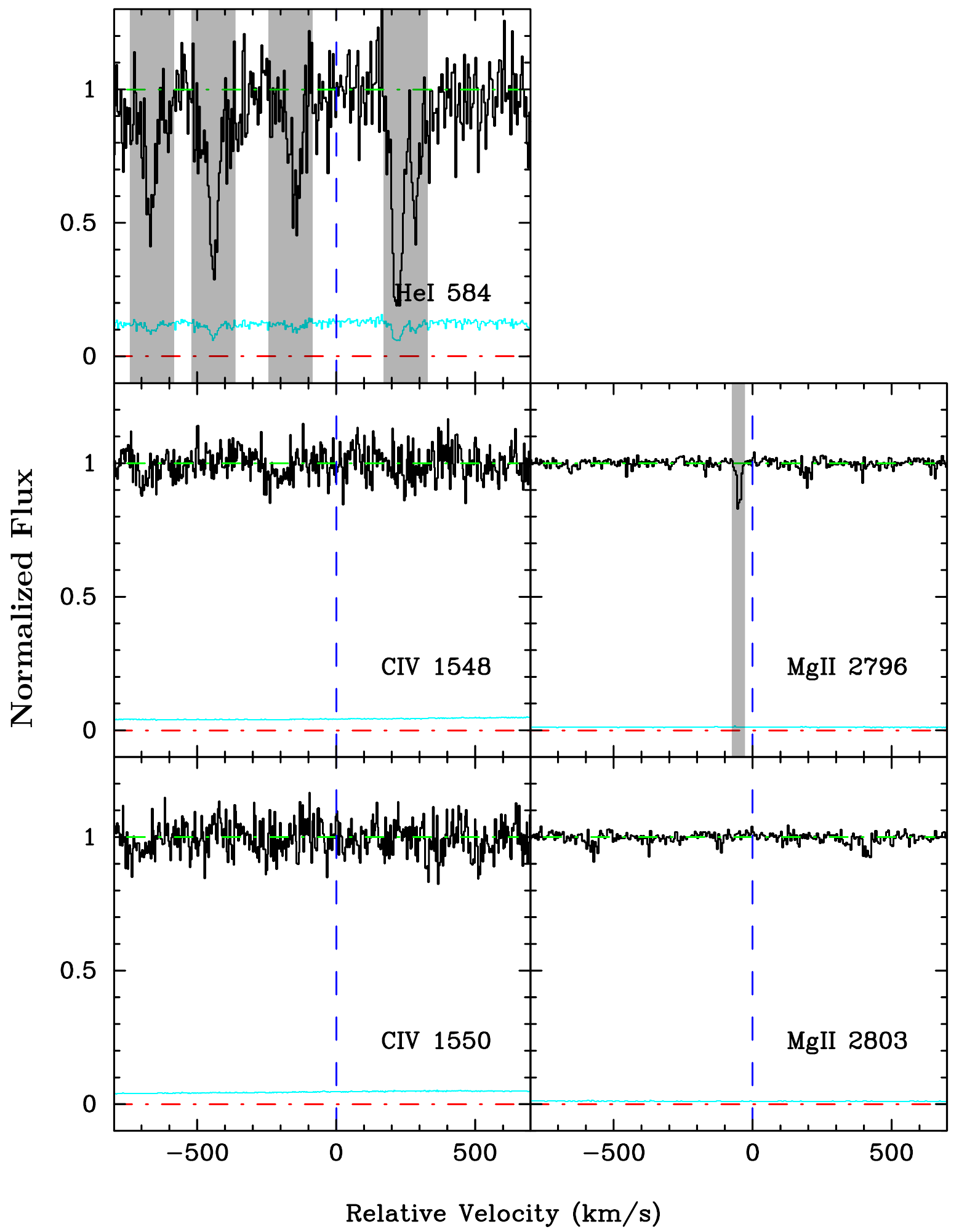}  
}} 
}}  
\vskip+0.0cm  
\caption{
Constraints on the UV absorption features for a subset of strong transitions commonly seen in quasar absorption-line surveys.  Zero velocity corresponds to the quasar emission redshift \zem = 1.3531.  Contaminating features associated with systems in the foreground at $z<$\,\zem\ are greyed out. At this redshift, \hi\ Lyman series lines are redshifted out of the COS FUV spectral coverage.  Therefore, constraints for neutral gas are based on \ion{He}{i}.  The available UV and optical absorption spectra show that no absorption features are detected in the vicinity of the 21cm absorption to within sensitive column density limits (see Table \ref{tab:uvlines}).  } 
\label{fig:uvopt}   
\end{figure}

Despite the absorption detected in \hi\ 21-cm, no absorption transitions in the UV are seen in available FUV and optical spectra of the quasar.  As described in Section~\ref{sec:mikeobs}, available HST COS and Magellan MIKE spectra provide spectral coverage for \ion{He}{i}, \ion{C}{ii}, \ion{C}{iv}, \ion{O}{iii}, \ion{Al}{ii}, \ion{Al}{iii}, \ion{Mg}{ii}, \ion{Fe}{ii} and \ion{Ca}{ii} at the redshift of the 21-cm absorber. In Fig~\ref{fig:uvopt}, we show FUV / optical spectra at the locations of some of these lines. The lack of both low and intermediate ionization species indicates that the absence of signals is unlikely due to the ionization state of the gas. Further, the lack of \ion{He}{i}, a non-metal, indicates that the absence of cold gas signatures cannot be attributed to possibly a low metal content.  While the data do not provide coverage for the hydrogen Lyman series lines, the observed 2$\sigma$ upper limit of \ion{He}{i} can be converted to an \hi\ column density limit of $\log\,N({\rm \hi\,})/{\rm cm}^{-2}<15.2$ \citep[e.g.,][]{Qu:2023}.  In Fig.~\ref{fig:uvopt} and Table~\ref{tab:uvlines}, we present the constraints imposed by \ion{C}{iv}, \ion{Mg}{ii}, and \ion{He}{i}, the strongest UV transitions available at this redshift.

It is worth mentioning here that in the FUV band ($\lambda_{eff}\sim 1528$~\AA) of the Galaxy Evolution Explorer \citep[GALEX;][]{Bianchi2017} observations, the quasar is detected with an AB magnitude of 17.91 mag. This detection directly rules out the presence of a DLA, or even a Lyman Limit System (LLS), at z=1.3531 along the UV line-of-sight, independent of any assumptions about metallicity or ionization state. This is because the Lyman limit at $z=1.3531$ occurs at a much redder wavelength ($\lambda\sim 2146$~\AA) than the cut-off of the GALEX FUV filter. If there were such a strong \hi\ absorber toward the sightline, GALEX would not detect the quasar in FUV images.

\begin{table}
\caption{UV absorption line constraints for G0 and G1 fitted to the 21-cm line. }
\vspace{-0.4cm}
\begin{center}
\begin{tabular}{cccccc}
\hline
\hline
{\large \strut}    Species     &  & \multicolumn{2}{c}{$\log\,N_{\rm X}$/(\cmsq)}  \\
\cline{3-4}
        X   & & G0 &  G1  \\
\hline
  \civ\   & & $<$12.5 & $<$12.7  \\
  \mgii\  & & $<$11.5 & $<$11.5  \\
  He~{\sc i} & & $<$13.6 & $<$13.4  \\
\hline

\end{tabular}
\label{tab:uvlines}
\end{center}
\end{table}

\subsection{Origin of the absorbing gas}
\label{sec:origin}

\begin{figure*} 
\includegraphics[
trim = {0cm 0cm 0cm 0cm}, clip=true,
width=0.95\textwidth,angle=0]{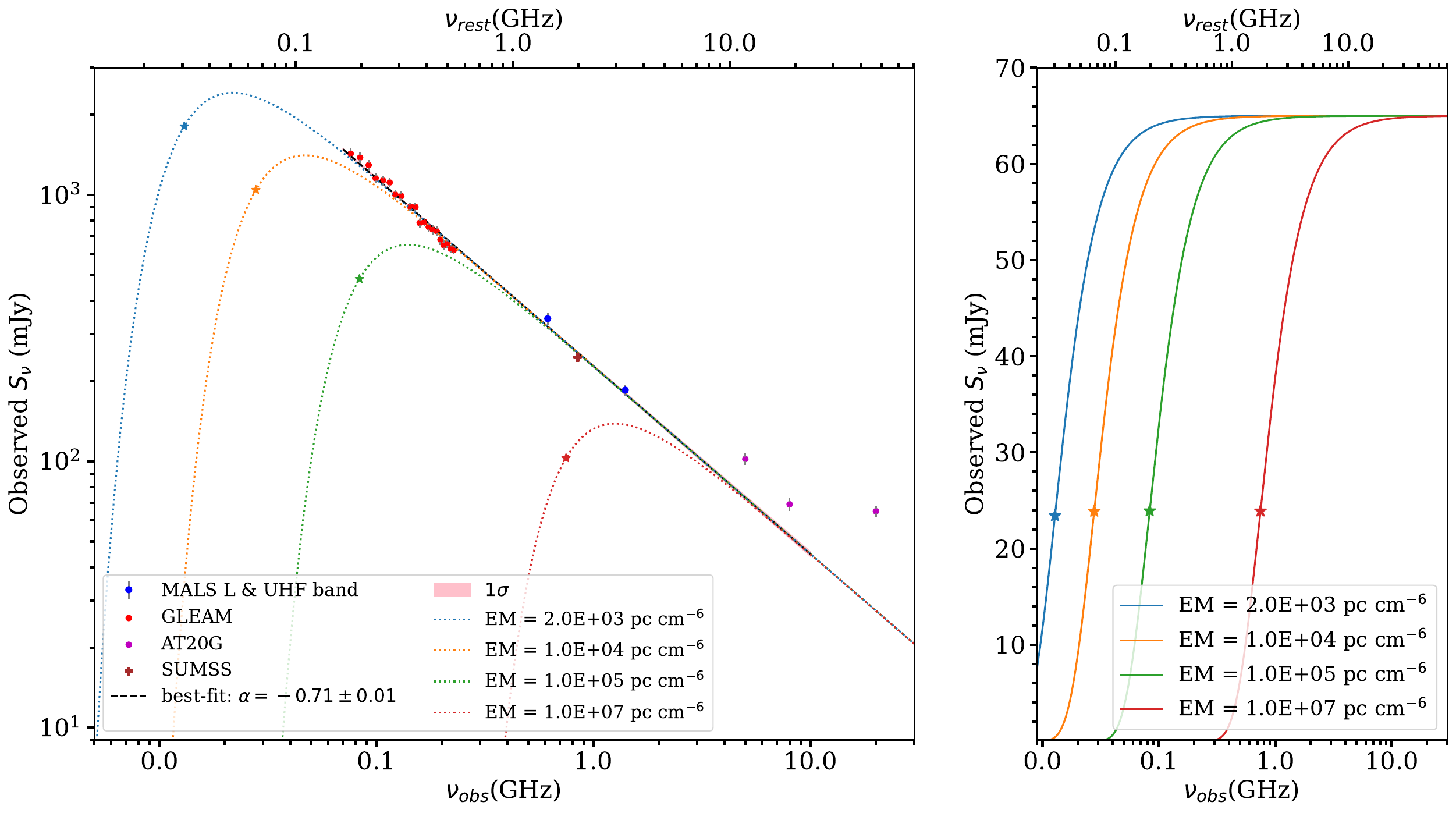}  
\vskip+0.0cm  
\caption{Left panel: The spectral energy distribution of J2339-5523 obtained by combining measurements from GLEAM (red circles), MALS L- and UHF-bands (blue circles), SUMSS (brown `+') and AT20G catalog (magenta circles). A power law ($S_\nu \propto \nu^{\alpha}$) fit (black dashed line) to MALS and GLEAM flux densities yields a spectral index, $\alpha = -0.71\pm0.01$. The dotted lines represent SED attenuated by FFA due to a range of emission measures. Right panel: Attenuation of flat-spectrum `core' component due to FFA for a range of emission measures. In both the panels,  asterisks represent $\nu_p$ (refer to Section~\ref{sec:origin}).
} 
\label{fig:radsedffa}   
\end{figure*}

The CSS sources such as J2339-5523 typically exhibit a core-jet or a double-lobed structure reminiscent of large scale radio sources \citep[][]{Odea21}.  Since J2339-5523 is a quasar, we anticipate a dominant `core' component.  Indeed,  near-simultaneous high-frequency radio observations at 5, 8 and 20\,GHz from AT20G survey carried out using the Australia Telescope Compact Array (ATCA) over 2004 - 2008 \citep[][]{Murphy10} exhibit a flat spectrum component ($\alpha_{\rm ATCA}$ = -0.3$\pm$0.2) with flux densities of 102$\pm$5 , 69$\pm$4 and 65$\pm$3 \,mJy, respectively.  
We note that in the W3 (12$\mu m$) and W4 (24$\mu m$) bands of the Wide-field Infrared Survey Explorer \citep[WISE; ][]{Wright10, Cutri14}, the quasar has a brightness of 8.475 $\pm$ 0.022\,mag and 6.033 $\pm$ 0.047\,mag, respectively.  For non-AGN sources the 33\,GHz flux density (S$_{33\,GHz}$) in mJy is related to flux density at 24\,$\mu m$ ($f_{24\mu m}$) in mJy through $f_{24\mu m}$ = 140$\times$S$_{33\,GHz}^{0.94}$ \citep[][]{Murphy12}. The W3 and W4 band flux densities correspond to rest frame fluxes at $\sim$5$\mu m$ and $\sim$10$\mu m$, respectively. Assuming a simple power law holds in the range (5--24$\mu m$), we derived the rest frame 24$\mu m$ flux density to be $\sim 235$\,mJy, corresponding to S$_{33\,GHz}$ $\approx 1.73$~mJy. Using this value, we found that the mid-infrared flux density, if due to star formation, will contribute only about $2.7$\% to the optically thin emission detected in AT20G measurements (rest frequency $\sim$ 10 - 40\,GHz and core flux density $\sim 65$~mJy). 
Hereafter, we will consider the entire flux density of this quasar as measured in the AT20G survey to be associated with the AGN i.e., the radio `core'. 

The detected \hi\ 21-cm absorption may stem from gas within the circumnuclear disk, or it could arise from kinematically disturbed gas affected by AGN / stellar feedback, or may have its origin in the regularly rotating gas within the galactic disk. Depending on the location, extent and clumpiness of the gas, it may only partially cover the radio emission.  
Due to its central location, the detection of absorption towards the radio `core' or absence of it can provide crucial constraints on the origin of the absorbing gas \citep[see e.g.,][]{Carilli98, Gupta22m1540}.  In the case of J2339-5523, the contribution of the radio `core' to the detected 21-cm absorption will primarily depend on its brightness near the 21-cm line frequency and the actual column density of the gas.  
For a given column density and covering factor, the strength of the \hi\ 21-cm absorption towards the `core' can be suppressed in the following two ways: (i) suppression of the background radio continuum due to free-free absorption (FFA) or synchrotron self-absorption (SSA), or (ii) excitation mechanisms raising the spin temperature of the gas, especially due to the proximity to the AGN.

For the modification of radio SED, both FFA and SSA may provide viable solutions. But in the latter case, the models depend strongly on the flux density and the frequency of the low-frequency turnover, as well as the size of the radio sources \citep[e.g.,][]{Orienti12, Callingham15, Shao22}.  Since we do not have observational constraints on either of these, we consider the viability of a simple homogeneous FFA scenario \citep[][]{Bicknell97}.  For this we will assume that the determination of radio SED is not affected by temporal variability.  This is a reasonable assumption considering that a single power-law model can provide a reasonable fit ($\alpha$ = -0.71 $\pm$ 0.01) to GLEAM, SUMSS, MALS L- and UHF-band measurements, spread over more than 20 years (see left panel of Fig.~\ref{fig:radsedffa}).   
%

The attenuation, i.e., optical depth, due to FFA is related to electron temperature ($T_e$) in Kelvin, frequency ($\nu$) in GHz and free electron density ($n_e$) in cm$^{-3}$ through 
\begin{equation}
    \label{eq:ffa}
    \tau_\nu = 8.24\times10^{-2}  \Big[\frac{T_e}{\text{K}}\Big]^{-1.35} \times \Big[\frac{\nu}{\text{GHz}}\Big]^{-2.1} \times \Big[\frac{\text{EM}}{\text{pc cm}^{-6}}\Big],
\end{equation}
where EM = $\int n_e^2 dl$ is the emission measure and $l$ is the distance through the screen in pc \citep[][]{Mezger67}. 
In the left panel of Fig.~\ref{fig:radsedffa}, we also show the observed SED attenuated by the FFA optical depth, $\tau_\nu$ = ($\nu$/$\nu_p$)$^{-2.1}$, where $\nu_p$ is the frequency at which FFA optical depth is 1.  We assume $T_e$ = $10^4$\,K. At the lowest frequency measurements from GLEAM, no signature of flattening in the radio SED  is observed. This implies that any external screen covering the radio source has emission measure of $<2\times10^{3}$\,pc\,cm$^{-6}$ (see blue dotted curve in the left panel of Fig.~\ref{fig:radsedffa}).
%

However, it is possible that the radio `core' is embedded in a denser environment.  In the right panel of Fig.~\ref{fig:radsedffa}, we show the attenuation of the radio core's flux density due to FFA.  Clearly, an external ionized screen with emission measure (EM$\sim10^7$\,pc\,cm$^{-6}$) will be required to attenuate the continuum flux density of the flat-spectrum core component at the redshifted \hi\ 21-cm line frequency to a few mJy such that it will not produce detectable 21-cm absorption. The unresolved emission from the source does not allow us to investigate the scenario of the core getting attenuated by FFA any further. Hence, while the suppression due to FFA cannot be completely ruled out, the non-detection ($\log\,N({\rm \text{\ion{He}{i}}})/{\rm cm}^{-2}<13.8$) of associated UV absorption signatures in the optical spectrum (Fig.~\ref{fig:uvopt}) point towards the simplest scenario where no large \hi\ column ($N$(\hi)$>10^{17}$\,\cmsq) is present towards the radio `core' and optical AGN, regardless of whether the `core' is attenuated by FFA or not. The optical/UV observations also rule out the possibility of a high-$T_{\rm s}$ \hi\ gas in front of the radio `core'. In the absence of any direct constraints on the morphology of the radio source, hereafter, we will assume it to possess the canonical structure comprising a radio `core' with lobes on its either side, being viewed close to the jet axis.

\section{Properties of absorbing gas: is J2339-5523 unique?}
\label{sec:QSOvsGAL}

\begin{figure}[t]
\centerline{\vbox{
\centerline{\hbox{ 
\includegraphics[
trim = {0cm 0cm 0cm 0cm}, clip=true,
width=0.50\textwidth,angle=0]{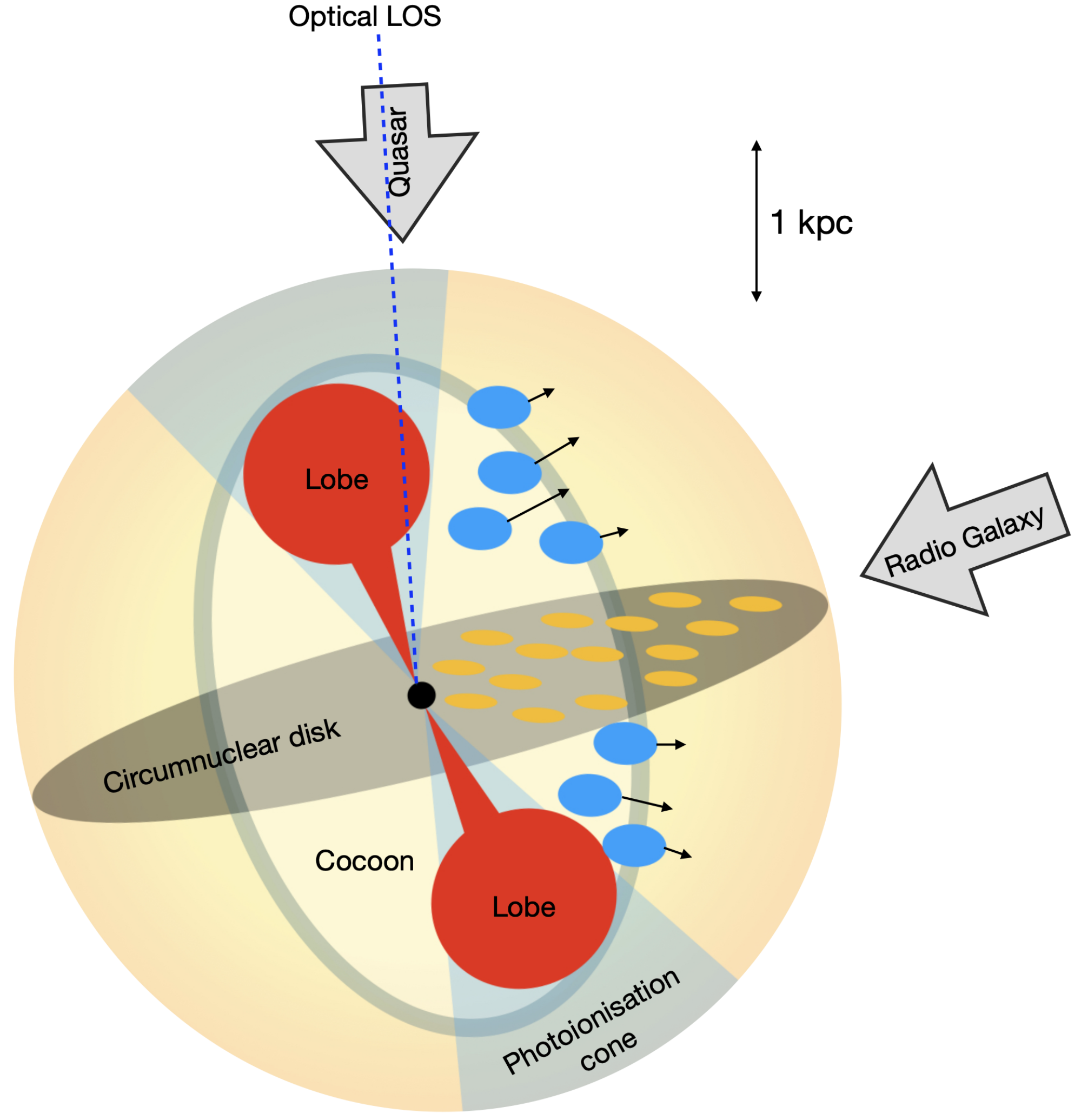}  
}} 
}}  
\vskip+0.0cm   
\caption{Cartoon depicting quasar and radio galaxy views of an AGN. \hi\ clouds in yellow are located in the circumnuclear disk, whereas clouds in blue are offset from the disk plane.  The optical line of sight is indicated by a blue dashed line, sampling little amount of \hi\ clouds. The sightlines to quasars primarily pass through photoionization cone and sample cold gas closer to the jet-axis and inner circumnuclear disk. The receding lobe would sample both positive and negative velocities corresponding to blue clouds.  The galaxy view, with weak core, primarily samples cold gas from blue clouds (negative velocities) in front of lobes and the galaxy-wide ISM outside the photoionization cone. 
}
\label{fig:cartoon}
\end{figure}

\begin{figure} 
\centerline{\vbox{
\centerline{\hbox{ 
\includegraphics[trim = {0cm 0cm 0cm 0cm}, clip=true, width=0.50\textwidth,angle=0]{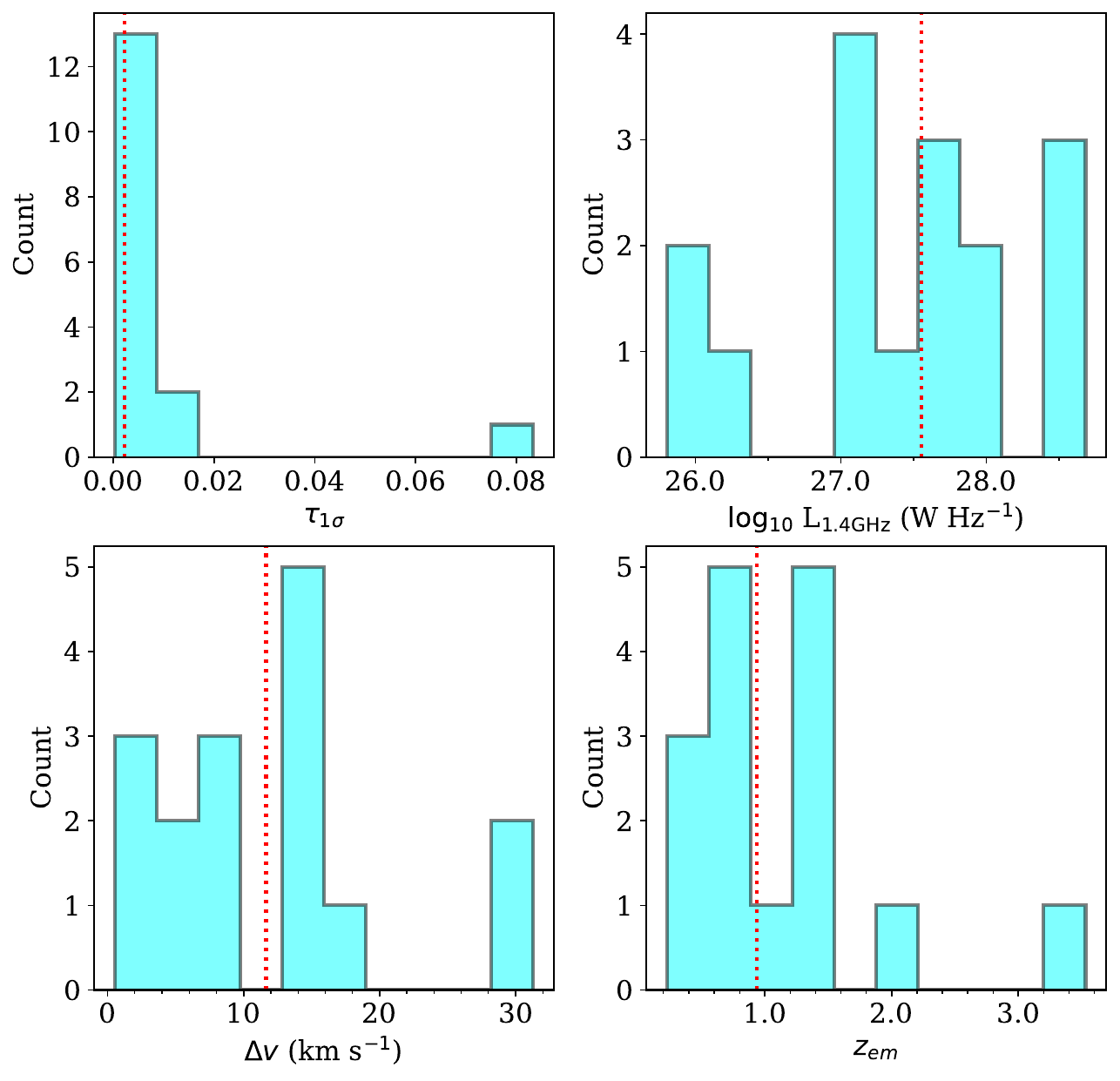}  
}}
}}
\vskip+0.0cm  
\caption{
Distribution of $\tau_{1\sigma}$ (top left), radio luminosity at 1.4~GHz (L$_{1.4\rm GHz}$ in W\,Hz$^{-1}$; top right), spectral resolution ($\Delta v$ in \kms; bottom left) and redshift ($z_{em}$; bottom right) for the sample of 16 quasars (including J2339-5523) with associated \hi\ 21-cm detections. The vertical dotted lines indicate the median values of each parameter on the $x$-axis.}
\label{fig:sample_dist}   
\end{figure}


The properties of \hi\ 21-cm absorption associated with radio sources are known to be related to its orientation and size. Among compact radio sources ($<$15\,kpc), the detection rates are found to be significantly lower for quasars than radio galaxies \citep[][]{Gupta06uni}.  This is consistent with the unification scheme implying that lines of sight to the radio components for quasars more frequently pass through the ionization cone which is in general devoid of cold gas (see Fig.~\ref{fig:cartoon}), although, trace molecular gas have been detected through these channels as well \citep{Noterdaeme2021}.  
In fact, the jet-axis is not always perfectly aligned with respect to the circumnuclear disk \citep[][]{Ruffa20}, and a jet-ISM interaction may also push the gas over a wider range of angles forming a cocoon, thereby increasing its covering factor for absorption \citep[][]{Wagner12, Mukherjee16}. 
Additionally such jet-ISM interaction can also be responsible for the development of the strong large-scale outflows, which can entrain a large amount of the gas and facilitate the formation of the cold gas by condensation of the medium at the envelope of the shocked gas \citep[e.g.,][]{Villar-Martin1999, Morganti17}.
These factors along with the intrinsic misalignment in the radio morphology may be responsible for producing \hi\ 21-cm absorption in some of these cases \citep[e.g.,][]{Gupta22m1540}.

At optical wavelengths, the detection of H\,{\sc i} clouds can be tricky due to leaking Ly-$\alpha$ emission \citep[e.g.][]{Finley2013,Fathivavsari2018}, but the presence of cold gas in the quasar environment can be revealed by H$_2$ absorption towards the central engine \citep[][]{Noterdaeme2019}. In these cases, large molecular reservoirs likely in the host galaxy are also frequently seen from CO emission \citep[e.g.,][]{Dasyra2012, Noterdaeme2023}.
At redshifted 21-cm line frequencies, the prominent jets and lobes along with the radio `core' provide multiple sight lines to more efficiently sample the distribution and kinematics of gas in the circumnuclear disk and host galaxy ISM (see Fig.~\ref{fig:cartoon}). 

In the literature, \hi\ 21-cm absorption detection associated with 15 quasars have been reported (see Table~\ref{tab:litqso}).  These form a very heterogeneous sample spanning a wide range in redshift (\zem = 0.2 - 3.5) and 1.4\,GHz spectral luminosity (L$_{1.4\rm GHz}$ = $10^{26 - 29}$\,W\,Hz$^{-1}$). These have been observed with varied spectral resolution in terms of $\Delta v$  ( 0.54 - 31.3 \kms\,) and optical depth sensitivity ($\tau_{1\sigma}$).  These basic properties are listed in Table~\ref{tab:litqso} and their distributions are shown in Fig.~\ref{fig:sample_dist}.  We smooth all the spectra to a common velocity resolution of $\sim$30~\kms\, and measure the following two quantities as proxies for gas kinematics from the absorption profiles: 
(i)  $\Delta V_{\rm null}$, i.e., width between the nulls of the absorption profile (in \kms\,) indicating the velocity spread of the absorbing gas; and, (ii) $A_p = |V_{\rm null, B}| - |V_{\rm null, R}|$ to quantify the blueward or redward asymmetry in the line profile. To measure $\Delta V_{\rm null}$, we fit the smoothed absorption profile using multiple Gaussians and determine nulls as velocities where the fitted profile falls below $1\sigma$ optical depth sensitivity.  In Table~\ref{tab:litqso}, we provide blueward ($V_{\rm null, B}$) and redward ($V_{\rm null, R}$) null velocities with respect to the 21-cm absorption peak. From these $\Delta V_{\rm null}$ can be obtained using $\Delta V_{\rm null} = |V_{\rm null, B} - V_{\rm null, R}|$. For J2339$-$5523, $\Delta V_{\rm null}$ = 397\,\kms\ is typical of the quasar sample.  However, the profile is slightly redshifted with absorption peak at $\sim$200\,\kms\ (Fig.~\ref{fig:21cm}). The absorption component `G$_1$' is consistent with the systemic velocity and may represent clouds in the circumnuclear disk. The redshifted component `G$_0$' may then correspond to inflowing gas.  But this simplistic interpretation is compromised by the simultaneous possibility of multiple origins of the gas \citep[][]{Combes23}.

\begin{figure*} 
\centerline{\vbox{
\centerline{\hbox{ 
\includegraphics[trim = {0cm 0cm 0cm 0cm}, clip=true, width=\textwidth,angle=0]{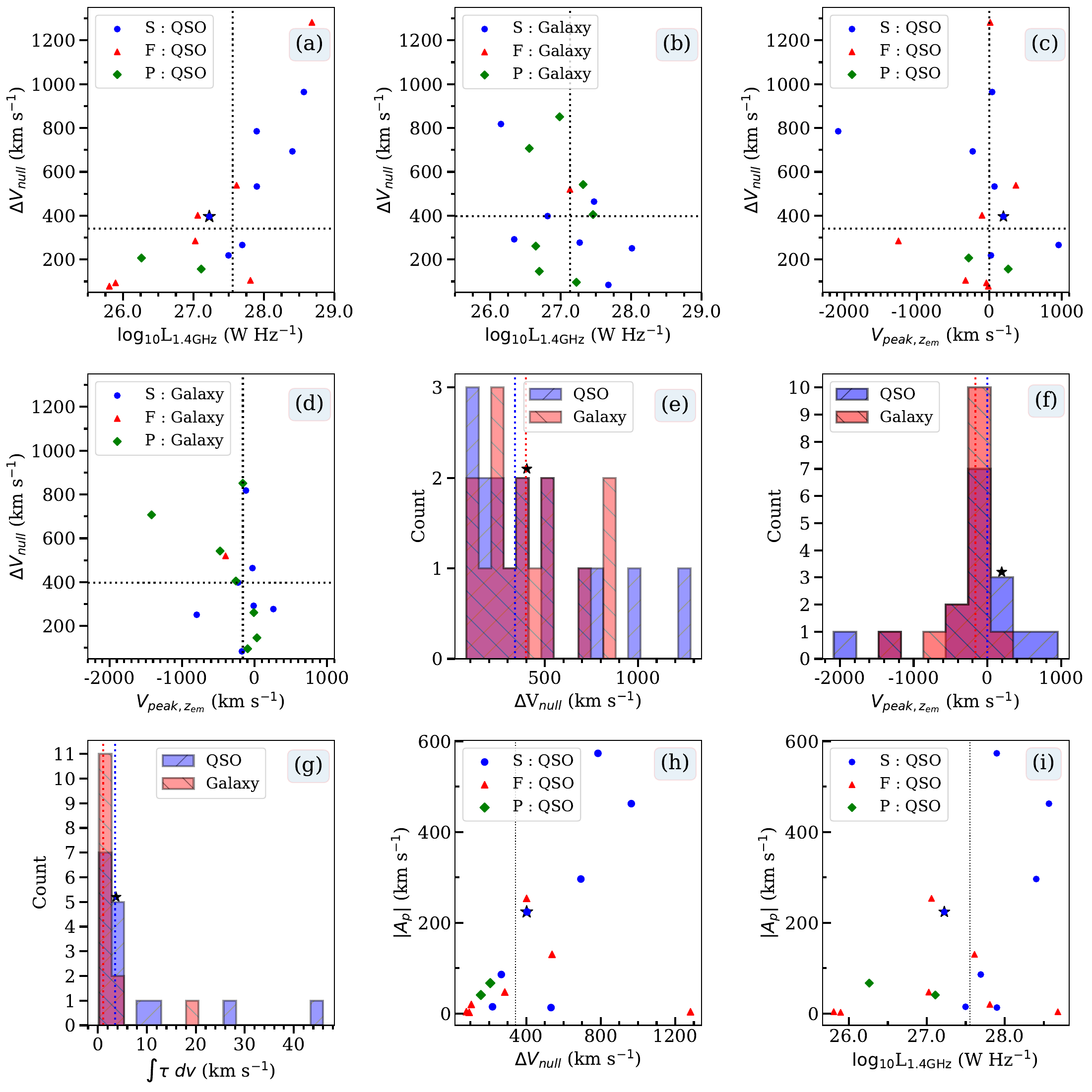}  
}}
}}
\vskip+0.0cm  
\caption{
Distributions of various observables associated with \hi\ 21-cm absorption line and radio continuum associated with quasars (16) and radio galaxies (19). 
    Different symbols and colors are used for systems with different radio SEDs: steep spectrum (S, filled blue circles), peaked spectrum (P, filled green diamonds) and flat spectrum (F, filled red triangles). 
    In all the panels, dotted lines mark the median and the location of J2339-5523, a steep spectrum source, is shown using $\star$. }
\label{fig:correlations}   
\end{figure*}


We investigate the origin of the absorbing gas by comparing the \hi\ absorption properties of the quasar sample with a sample of 19 radio-galaxies from the literature \citep[][]{Vermeulen03, Aditya2019}, matched in 1.4\,GHz luminosity with the quasar sample and with associated \hi\ 21-cm absorption detection. The details are provided in Table~\ref{tab:litgals}. 
For the quasar sample, we find a significant correlation between $L_{1.4\rm GHz}$ and $\Delta V_{\rm null}$ (Kendall's tau coefficient, r$=0.65$ and $p-$value = $2\times 10^{-4}$).  The corresponding scatter plot is shown in Fig.~\ref{fig:correlations}(a). Interestingly, $\Delta V_{\rm null}$ for radio galaxies  exhibit no correlation (r = $-0.08$; p-value = $0.68$; Fig.~\ref{fig:correlations}(b)).  However, the extent of absorption lines are similar to the quasar sample (Figs.~\ref{fig:correlations}(c)-(f)).  The galaxies show  marginally larger $\Delta V_{\rm null}$ (KS test $p$-value = 0.70) and blueshifts (KS test $p$-value =  0.37) measured with respect to optical redshifts. Also, we do not find any significant correlations between $|A_p|$ and $L_{1.4\rm GHz}$ for quasars and galaxies. However, between $|A_p|$ and $\Delta V_{\rm null}$ a moderate correlation in case of quasars and a relatively stronger correlation in case of galaxies is observed.
The Kendall's tau-coefficients and $p$-values for quasars and galaxies are 0.48 and 0.008, and 0.60 and 0.0001, respectively. This suggests that broader lines associated with powerful AGN (L$_{1.4\rm GHz}$ > $10^{26}$\,W\,Hz$^{-1}$) may also be more asymmetric, as seen among relatively less powerful AGN \citep[L$_{1.4\rm GHz}$ = $10^{24-26}$\,W\,Hz$^{-1}$;][]{Maccagni17}.

The quasars and radio galaxies with different radio SEDs i.e., steep spectrum (S; $\alpha\leq -0.5$), peaked spectrum (P; turnover at GHz frequencies) and flat spectrum (F; $\alpha > -0.5$) are also distinguished in Fig.~\ref{fig:correlations}.  The spectral classifications provide important information on both the size and nature of the background radio emission. Radio sources with higher turnover frequencies are more compact and vice-versa,  flatter spectra indicate a dominant core component, while emission from older population of electrons (e.g., lobes) gives rise to steeper spectra.
Generally, core-dominated and peaked spectrum sources have smaller projected largest linear sizes (LLS) estimated from milliarcsecond scale images (see Tables~\ref{tab:litqso} and ~\ref{tab:litgals}).  
However, note that LLS have been estimated using high-frequency ($>$2.3\,GHz) milliarcsecond scale images and present only a very approximate picture of radio emission close to the \hi\ 21-cm line frequencies.
Nevertheless, we find no significant correlation between LLS and $\Delta V_{\rm null}$ for the quasar sample (r = 0.27 ; $p$-value = 0.17), indicating that the observed $L_{1.4\rm GHz}$ and $\Delta V_{\rm null}$ relationship is not influenced by other observables such as the LLS \citep[see Fig.~9 of][]{Gupta12}.

\begin{figure}[t]
\centerline{\vbox{
\centerline{\hbox{ 
\includegraphics[
trim = {0cm 0cm 0cm 0cm}, clip=true,
width=0.40\textwidth,angle=0]{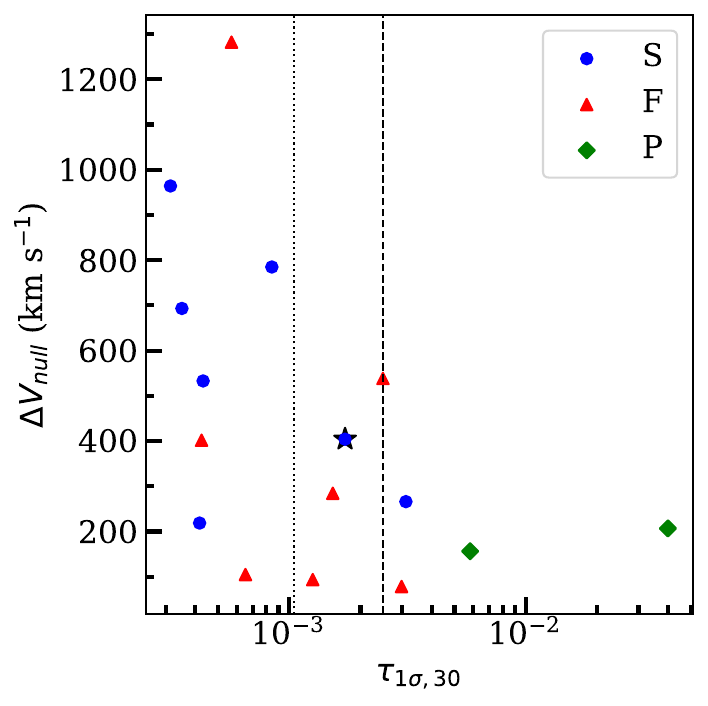}  
}} 
}}  
\vskip+0.0cm   
\caption{ Velocity width between the nulls of the absorption profiles, $\Delta V_{\rm null}$ (in \kms\,) as a function of $\tau_{1\sigma,30}$ for all the 16 quasars listed in Table~\ref{tab:litqso}. The markers and dotted vertical lines are same as in Fig~\ref{fig:correlations}. The dashed vertical line marks the median $\tau_{1\sigma,30}$ of higher $\tau_{1\sigma,30}$ bin.}
\label{fig:biases}
\end{figure}

The luminosity (L$_{1.4\rm GHz}$) and redshift are strongly correlated in the quasar sample. Varying optical depth sensitivity, either due to selection of brighter AGNs or poor spectral rms, across a given sample may also affect the measurements of $\Delta V_{\rm null}$ and $A_p$. A Kendall's tau test for the quasars does indicate a weak anti-correlation, with statistical significance at less than 2$\sigma$ (tau $=-0.35$, $p-$value = $0.07$). This suggests that sensitivity issues due to Malmquist bias may have minimal impact on our results, as illustrated in Fig.~\ref{fig:biases}.
We note that 14 out of 16 systems ($\sim$ 88\%) in the quasar sample exhibit optical depth sensitivities within 1$\sigma$ of the median sensitivity (Fig.~\ref{fig:sample_dist}). 
We investigate the potential biases due to optical depth sensitivity by dividing the quasar sample into two equal groups at median $\tau_{1\sigma,30}$. We then up-scaled the rms noise in the spectra with better sensitivities to match the common median $\tau_{1\sigma,30}$ of the lower sensitivity subset. 
This rendered absorption profiles of 5 systems unsuitable for the analysis.  The remaining 11 systems still exhibit the correlation between $\Delta V_{\rm null}$ and L$_{1.4\rm GHz}$ (tau $=0.56$, $p-$value = $0.01$). However, the weaker correlation between $\Delta V_{\rm null}$ and $|A_p|$ seen in Figs.~\ref{fig:correlations}(h) disappeared. 
We are unable to perform a similar detailed investigation for the sample of radio galaxies due to the unavailability of adequate information in the literature. However, since these are generally bright radio sources, we do not expect any additional biases, except due to any intrinsic redshift evolution among quasars ($z_{median}$ = 0.7) and radio galaxies ($z_{median}$ = 0.4).

Overall, the correlation of L$_{1.4\rm GHz}$ and $\Delta V_{\rm null}$ among powerful quasars but not radio galaxies suggest that the orientation of AGN may play an important role in observed gas properties. As illustrated in Fig.~\ref{fig:cartoon}, in the radio-galaxy view the sight lines towards radio lobes would be effective in tracing the kinematically disturbed gas within the cocoon created by the jet-ISM interaction and galaxy-wide ISM outside the photoionization cone.
In the quasar view, the jet-axis is closely aligned with the line-of-sight. Therefore, a significant portion of the absorbing clouds, particularly at extreme velocities, may (scenario-1) originate from the inner circumnuclear disk experiencing strong gravitational field of the central black hole  or (scenario-2) perturbed by energy output associated with radio jets. Since powerful AGN are associated with massive red galaxies \citep[e.g.,][]{Goulding2014}, the role of starbursts in gas perturbations may be ignored. The conclusion is also supported by the mid-infrared colors of AGN considered here \citep[see][]{Wright10}.  
In-line with scenario-1, the integrated optical depths towards quasars probing inner regions of the galaxy with higher filling factor of cold gas are also marginally larger (Fig.~\ref{fig:correlations}(g); KS test $p$-value = 0.05). Fig~\ref{fig:cartoon} illustrates that in the quasar view (scenario-2), the sight line towards the receding lobe may trace both inflowing and outflowing clouds ($<$1\,kpc for the sample) exhibiting positive and negative velocities, respectively, thereby resulting in symmetric line profiles. This may explain the less asymmetric absorption lines observed in the case of quasars, and broader/ more  blueshifted absorption profiles observed among radio-galaxies. Note that scenario-1 assumes that more powerful quasars at higher redshifts in the sample are associated with more massive black holes.  While this can not be ruled out \citep{Labita2009b}, a larger complete sample observed with uniform sensitivity is required to draw conclusions.

Interestingly, despite the incompleteness of the sample, and non-uniform quality and wavelength coverage of the radio spectra, the case of J2339-5523 does not appear unique.  In fact, in all of the above it is interesting to note that J2339-5523 exhibits typical properties among quasars with \hi\ 21-cm absorption (see $\star$ in Fig.~\ref{fig:correlations}).

Finally, we examine the generality of the discrepancy between radio and optical sight lines in the case of J2339-5523 using the samples in Tables~\ref{tab:litqso} and \ref{tab:litgals}. 
Optical spectra are available for 5 quasars (see Table~\ref{tab:EW}).  In only one case, absorption lines of Mg~{\sc ii} and Mg~{\sc i}, indicators of high-$N$(\hi) gas, are detected.  Among the remaining four cases no absorption is detected.  Even in the case of J080133.56$+$141443.0, although strong Mg~{\sc ii} doublets as well as Mg~{\sc i} absorption lines are detected in the SDSS DR6  spectrum \citep{SDSSDR6}, the radio core is self-absorbed at the frequency of the \hi\, 21-cm absorption \citep{Ishwara03}. Hence, towards J0801+1414 the cold gas seen in 21-cm absorption is not against the radio core and does not correspond to the same gas producing metal absorption in the optical spectrum.

Optical spectra adequate to search for absorption lines are available for 7/19 radio galaxies (see Table~\ref{tab:EW})
In all the cases, signatures of cold gas through absorption lines of Ca~{\sc ii}\,$\lambda\lambda$3934,3969 and / or Na~{\sc i}\,$\lambda\lambda$5891,5897 are reliably detected. The peak absorption in 5/6 cases also matches well with the \hi\ 21-cm absorption. 
The increased occurrence of concurrent cold gas detection in both radio and optical observations for radio galaxies suggests both are tracing same population of clouds, in the less clumpy outer regions of the galaxy.

\section{Summary}
\label{sec:summary}

We have reported the detection of \hi\ 21-cm absorption associated with the quasar J2339$-$5523 at $z=1.3531$. Such detections are rare and only six at $1<z<2$ are known. The detected absorption profile is wide with a velocity width between nulls, $\Delta V_{\rm null}$ = 397\,\kms. However, the profile is slightly redshifted with absorption peak at $\sim$200\,\kms\ (Fig.~\ref{fig:21cm}). 
Assuming $T_s = 100$ K and $f_c = 1$, we derived a lower limit to $N$(\hi\,) $\geq$ $6.7\times 10^{20}$ \cmsq\,. Interestingly, the FUV and optical spectra towards this source obtained using HST/COS and MIKE spectrographs, respectively, do not reveal any signature of metal absorption.  The simplest explanation would be that no large \hi\ column ($N$(\hi) $>10^{17}$\,\cmsq) is present towards the radio `core' and the optical AGN.

%
Based on the joint optical and radio analysis of a heterogeneous sample of 16 quasars ($z_{median}$ = 0.7) and 19 radio galaxies ($z_{median}$ = 0.4) with \hi\ 21-cm absorption and matched in 1.4\,GHz luminosity (L$_{1.4\rm GHz}$ > $10^{26}$\,W\,Hz$^{-1}$), a consistent picture emerges where powerful quasars -- exhibiting a correlation between L$_{1.4\,\rm GHz}$ and $\Delta V_{\rm null}$ -- are  tracing the gas in the inner circumnuclear disk and / or clouds close to the jet-axis perturbed by the jet-ISM interaction.  Consistent with this, the integrated optical depths towards quasars probing inner regions of the galaxy with higher filling factor of cold gas are also marginally larger.  In contrast, the radio galaxies have absorption lines of similar extent but exhibit no correlation with the L$_{1.4\,\rm GHz}$. These absorbers most likely trace the gas from the extended cocoon created by the jet-ISM interaction and the galaxy-wide ISM outside the photoionization cone.  The absorption associated with these is also marginally broader, more asymmetric and blue-shifted with respect to the systemic velocity. While these data exhibit the different origins of absorbing gas detected in quasars and radio galaxies, a larger complete sample observed with uniform sensitivity is required for exact interpretation.

The \hi\ 21-cm absorption detection towards J2339-5523 and analysis presented here demonstrates the potential of radio spectroscopic observations, even when spatially unresolved, to reveal the origin of the absorbing gas through  multiple sight lines towards core, jet and lobes.  In the near future, a clearer picture of the origin and the structure of the absorbing gas will emerge from the larger samples from the ongoing surveys \citep[][]{Gupta17mals, Allison22, Hu23}. Specifically, in near future, bright radio sources from MALS will reveal the evolution of cold gas associated with a variety of AGN at $0<z<1.5$ \citep[][]{Deka24a}.


\begin{acknowledgements}
We thank an anonymous referee for useful comments.
The MeerKAT telescope is operated by the South African Radio Astronomy Observatory, which is a facility of the National Research Foundation, an agency of the Department of Science and Innovation. The MeerKAT data were processed using the MALS computing facility at IUCAA (\href{https://mals.iucaa.in/}{https://mals.iucaa.in/}).
EB acknowledges support by NASA under award number 80GSFC21M0002.
The National Radio Astronomy Observatory is a facility of the National Science Foundation operated under cooperative agreement by Associated Universities, Inc.  
This paper includes data gathered with the 6.5 meter Magellan Telescopes located at Las Campanas Observatory, Chile.
This paper has utilized data from the Cosmic Origins Spectrograph (COS) onboard HST.
This research has made use of NASA's Astrophysics Data System and the NASA/IPAC Extragalactic Database (NED), which is operated by the Jet Propulsion Laboratory, California Institute of Technology, under contract with the National Aeronautics and Space Administration.  

\end{acknowledgements}

\def\aj{AJ}%
\def\actaa{Acta Astron.}%
\def\araa{ARA\&A}%
\def\apj{ApJ}%
\def\apjl{ApJ}%
\def\apjs{ApJS}%
\def\ao{Appl.~Opt.}%
\def\apss{Ap\&SS}%
\def\aap{A\&A}%
\def\aapr{A\&A~Rev.}%
\def\aaps{A\&AS}%
\def\azh{AZh}%
\def\baas{BAAS}%
\def\bac{Bull. astr. Inst. Czechosl.}%
\def\caa{Chinese Astron. Astrophys.}%
\def\cjaa{Chinese J. Astron. Astrophys.}%
\def\icarus{Icarus}%
\def\jcap{J. Cosmology Astropart. Phys.}%
\def\jrasc{JRASC}%
\def\mnras{MNRAS}%
\def\memras{MmRAS}%
\def\na{New A}%
\def\nar{New A Rev.}%
\def\pasa{PASA}%
\def\pra{Phys.~Rev.~A}%
\def\prb{Phys.~Rev.~B}%
\def\prc{Phys.~Rev.~C}%
\def\prd{Phys.~Rev.~D}%
\def\pre{Phys.~Rev.~E}%
\def\prl{Phys.~Rev.~Lett.}%
\def\pasp{PASP}%
\def\pasj{PASJ}%
\def\qjras{QJRAS}%
\def\rmxaa{Rev. Mexicana Astron. Astrofis.}%
\def\skytel{S\&T}%
\def\solphys{Sol.~Phys.}%
\def\sovast{Soviet~Ast.}%
\def\ssr{Space~Sci.~Rev.}%
\def\zap{ZAp}%
\def\nat{Nature}%
\def\iaucirc{IAU~Circ.}%
\def\aplett{Astrophys.~Lett.}%
\def\apspr{Astrophys.~Space~Phys.~Res.}%
\def\bain{Bull.~Astron.~Inst.~Netherlands}%
\def\fcp{Fund.~Cosmic~Phys.}%
\def\gca{Geochim.~Cosmochim.~Acta}%
\def\grl{Geophys.~Res.~Lett.}%
\def\jcp{J.~Chem.~Phys.}%
\def\jgr{J.~Geophys.~Res.}%
\def\jqsrt{J.~Quant.~Spec.~Radiat.~Transf.}%
\def\memsai{Mem.~Soc.~Astron.~Italiana}%
\def\nphysa{Nucl.~Phys.~A}%
\def\physrep{Phys.~Rep.}%
\def\physscr{Phys.~Scr}%
\def\planss{Planet.~Space~Sci.}%
\def\procspie{Proc.~SPIE}%
\let\astap=\aap
\let\apjlett=\apjl
\let\apjsupp=\apjs
\let\applopt=\ao
\bibliographystyle{aa}
\bibliography{mybib}

\begin{appendix}

\section{List of quasars from the literature}
\label{sec:litqsos}

Here we present the properties of quasars selected from the literature. Table~\ref{tab:litqso} provides radio continuum and \hi\ 21-cm absorption line details.  The details from optical spectra are provided in Table~\ref{tab:EW}. 

\begin{sidewaystable*}
\tabcolsep=4pt
\caption{Quasars with \hi\ 21-cm absorption from literature. \label{tab:litqso}}
\centering
\begin{tabular}{lccccccccccllc}
\hline\hline             
Source name  & $z_{em}$(opt)  & $z_{abs}$ & V$_{null, B}$ & V$_{null, R}$ & S$_{1.4GHz}$ & $\alpha$ & $\log \rm L_{1.4 GHz}^*$ &  S$_{21-cm}$ & $\int \tau\ dv$   & 21-cm Ref. & LAS & LLS & Ref.\\
   &   &   & (\kms\,) & (\kms\,) & (mJy) &  & (\,W\,Hz$^{-1}$) &  (mJy) & (\kms\,)  &  & (arcsec) & (kpc) & \\
(1) & (2) & (3) & (4) & (5) & (6) & (7) & (8) & (9) & (10) & (11) & (12) & (13) & (14)\\
\hline
J000557.17$+$382015.2 & 0.229 & 0.229 & -45.4 & 48.2 & 572.7 & -0.03 $\pm$ 0.10 & 25.9 & 547.3 & 1.943 $\pm$ 0.057 & 1 & 0.043 & 0.162 & 1, 13 \\
J011137.31$+$390627.9 & 0.67 & 0.668 & -69.9 & 137.0 & 429.0 & 2.09 $\pm$ 0.05 & 26.3 & 180.0 & 46 $\pm$ 7 & 2 & 0.005 & 0.036 & 19\\
J050321.20$+$020304.6 & 0.5834 & 0.5848 & -98.7 & 57.7 & 2245.1 & 1.07 $\pm$ 0.04 & 27.1 & 1600.0 & 3.4 $\pm$ 0.4 & 2 & 0.011 & 0.073 & 19 \\
J061048.87$+$724853.2 & 3.52965 & 3.52983 & -643.0 & 639.0 & 1041.6 & -0.37 $\pm$ 0.07 & 28.7 & 1790.2 & 4.69 $\pm$ 0.15 & 3 & 0.050 & 0.373 & 3, 15\\
J080133.56$+$141443.0 & 1.1946 & 1.1929 & -198.4 & 494.9 & 2734.6 & -1.10 $\pm$ 0.11 & 28.4 & 5690.0 & 2.47$\pm$0.05 & 4 &  4.0 & 33.0 & 4\\
J090933.50$+$425346.5 & 0.67 & 0.67 & -259.8 & 273.5 & 4234.2 & -0.80 $\pm$ 0.10 & 27.9 & 6218.1 & 0.68 & 5,6 & 9.2 & 64.5 & 19 \\
J120321.94$+$041419.0 & 1.22429 & 1.20888 & -105.8 & 679.1 & 1146.6 & -0.65 $\pm$ 0.11 & 27.9 & 1675.2 & 2.52 $\pm$ 0.12 & 7 & 0.075 & 0.458 & 7, 16 \\
J124823.89$-$195918.8 & 1.275 & 1.275 & -713.3 & 250.8 & 5136.1 & -0.59 $\pm$ 0.04 & 28.6 & 8302.2 & 4.542 $\pm$ 0.043 & 7 & 0.040 & 0.230 & 20 \\
J145844.79$+$372021.5 & 0.33324 & 0.33316 & -41.3 & 37.0 & 215.1 & -0.02 $\pm$ 0.03 & 25.8 & 148.4 & 3.834 $\pm$ 0.079 & 1 & $<0.01$ & $<0.0495$ & 1, 14\\
J150609.53$+$373051.1 & 0.674 & 0.673 & -327.8 & 74.0 & 937.6 & 0.05 $\pm$ 0.10 & 27.1 & 1040.0 & 27.20 $\pm$ 0.04 & 8,9 &  0.055 & 0.387 & 9, 17\\
J154015.23$-$145341.5 & 2.104 & 2.1139 & -176.1 & 90.0 & 203.7 & -0.69 $\pm$ 0.04 & 27.7 & 652.0 & 11.30 $\pm$ 0.07 & 10 & $>$0.050  & $>$0.425 & 10, 18 \\
J154508.52$+$475154.6 & 1.277 & 1.2797$^{\dag}$ & -334.5 & 204.0 & 685.7 & -0.36 $\pm$ 0.09 & 27.6 & 861.5 & 9.56 $\pm$ 0.36 & 1 & 0.060 & 0.516 & 11\\
J181536.79$+$612711.6 & 0.601 & 0.594 & -118.6 & 166.0 & 849.3 & -0.48 $\pm$ 0.08 & 27.0 & 971.6 & 2.39 & 6 & 0.01 & 0.068 & 19 \\
J195542.73$+$513148.5 & 1.223 & 1.2206 & -62.5 & 42.3 & 1588.1 & 0.01 $\pm$ 0.09 & 27.8 & 1254.1 & 0.716 $\pm$ 0.037 & 12 & $>0.050^{\dag\dag}$ & $>0.4275^{\dag\dag}$ & 12\\
J233913.22$-$552350.4 & 1.353 & 1.355 & -86.4 & 310.8 & 184.2 & -0.71 $\pm$ 0.01 & 27.2 & 336.0 & 3.7 $\pm$ 0.2 & ... & $<$0.6 & $<5.2$ & ... \\
J225503.88$+$131333.9 & 0.543 & 0.543 & -101.8 & 116.9 & 2706.8 & -0.83 $\pm$ 0.11 & 27.5 & 3725.5 & 0.23 & 6 & 2.6 & 16.5 & 19\\
\hline
\hline
\end{tabular}
\tablebib{
(1) \cite{Aditya18}; (2) \cite{Carilli98}; (3) \cite{Aditya21}; (4) \cite{Ishwara03}; (5) \cite{Pihlstrom99}; (6) \cite{Vermeulen03}; (7) \cite{Aditya18gps}; (8) \cite{Carilli92}; (9) \cite{Kanekar08}; (10) \cite{Gupta21hz}; (11) \cite{Malkin2018}; (12) \cite{Aditya17}; (13) \cite{Fey00}; (14) \cite{Helmboldt07}; (15) \cite{Britzen07}; (16) \cite{Liu07}; (17) \cite{Polatidis95}; (18) \cite{Gupta22m1540}; (19) \cite{Gupta06}; (20)\cite{Sokolovsky11}.
}
\tablefoot{
\small{
    Column\,1: Source name (J2000); Column\,2: Optical redshift; Column\,3: Peak \hi\ 21-cm absorption redshift; Column\,4 - 5: Velocity to the bluer and redder side of the peak, respectively, where the optical depth goes below the spectrum sensitivity; Column\,6: Flux density at 1.4 GHz; Column\,7: Spectral index and its associated uncertainty; Column\,8: Radio luminosity at 1.4\,GHz; Column\,9: Flux density at the redshifted 21-cm line frequency; Column\,10: Total velocity integrated optical depth; Column\,11: \hi\ 21-cm absorption reference; Column\,12: Largest Angular Size in arcsec, and Column\,13: Largest Linear Size in kpc; Column\,14: Reference for LAS and LLS.\\}
    
    \footnotesize{$^{\dag}$ centroid of two Gaussian components of similar amplitudes\\
    $^{\dag\dag}$ size of the radio-core against which the absorption was observed (see \cite{Aditya17} for more details)\\
    $^*$ Derived using 1.4\,GHz (NVSS) and 4.85\,GHz flux densities from NED. In four cases, the spectral turnover falls between 1.4\,GHz and 4.85\,GHz, or 4.85\,GHz measurement is unavailable or dubious.  For these measurements at 1.4~GHz and at a lower frequency were used.\\}    
}
\end{sidewaystable*}

\begin{table*}[h]
\tabcolsep=4pt
\centering
\caption{Rest equivalent width (W$_r$ in $\AA$) estimates from optical spectra. \label{tab:EW}}
\begin{tabular}{lcccccc}
\hline \hline
Source name  & $z_{em}$(opt) &  Reference &$\lambda_{obs}$ ($\AA$) & $\lambda_{rest}$ ($\AA$) & Line & W$_r$ ($\AA$)\\
(1) & (2) & (3) & (4) & (5) & (6) & (7)\\
\hline
    \multicolumn{7}{c}{} \\
    \multicolumn{7}{c}{\large Quasars} \\
    \multicolumn{7}{c}{} \\
\hline
J0111$+$3906$^\dag$ & 0.67 & 1 & ... & 2694$-$5567 & MgII2796, MgI2853 & $<$4.633, $<$4.633 \\
J0801$+$1414 & 1.1946 & 2 & 3800$-$9200 & 1731$-$4192 &  MgII2796, MgI2853 & 1.865, 1.245\\
J0909$+$4253 & 0.67 & 2 & 3800$-$9200 & 2275$-$5509 &  MgII2796, MgI2853 & $<$0.744, $<$0.744 \\
J1203$+$0414 & 1.22429 & 2 & 3800$-$9200 & 1708$-$4136 & MgII2796, MgI2853 & $<$0.270, $<$0.266\\
J1458$+$3720 & 0.33324 & 2 & 3800$-$9200 & 2852$-$6900 & ... & ...\\
J1955$+$5131 & 1.223 & 3 & 5502$-$9841 & 2475$-$4427 & MgII2796, MgI2853 & $<$0.316, $<$0.139 \\
\hline
    \multicolumn{7}{c}{} \\
    \multicolumn{7}{c}{\large Radio galaxies } \\
    \multicolumn{7}{c}{} \\
\hline
J0834$+$5534 & 0.242   & 2 & 3800$-$9200 & 3059$-$7407 & NaI5891 & 2.0 \\
J0901$+$2901 & 0.194   & 2 & 3800$-$9200 & 3182$-$7705 & CaII3934, NaI5891 & 6.8, 1.5  \\
J1013$+$2448 & 0.94959 & 4 & 3800$-$9200 & 1949$-$4718 & CaII3934 & 1.2 \\
J1247$+$4900 & 0.20691 & 2 & 3800$-$9200 & 3148$-$7623 & CaII3934        & 6.0   \\
J1400$+$6210 & 0.431   & 2 & 3800$-$9200 & 2655$-$6429 & CaII3934, NaI5891 &  5.9, 2.7 \\
J1648$+$2224 & 0.82266 & 4 & 3800$-$9200 & 2085$-$5047 & CaII3934  & 7.0   \\
J2355$+$4950$^\dag$ & 0.2379  & 1 & ...     & 3036$-$7962 & CaII3934, NaI5891 & 7.8, 3.8 \\
\hline
\end{tabular}
\tablebib{(1) \cite{Lawrence96}; (2) \cite{SDSSDR6}; (3) \cite{Torrealba12}; (4) \cite{SDSSDR13}.}
\tablefoot{
\small{
    Column\,1: Source name (J2000); Column\,2: Optical redshift; Column\,3: Optical spectrum reference; Column\,4: $\lambda$ range (in $\AA$) covered in the observed frame; Column\,5: $\lambda$ range (in $\AA$) covered in the rest frame; Column\,6: Metal lines covered by the wavelength range; Column\,7: Rest equivalent width (W$_r$) estimates or upper limits.\\}
    
    \footnotesize{$^{\dag}$ the available optical spectrum is in the rest frame.}
}
\end{table*}

\section{List of radio galaxies from the literature}
\label{sec:litgals}

Here we present the properties of radio galaxies selected from the literature and matched in L$_{1.4\,\rm GHz}$ with the quasar sample. Table~\ref{tab:litgals} provides radio continuum and \hi\ 21-cm absorption line details.  The details from optical spectra are provided in Table~\ref{tab:EW}. 

\begin{sidewaystable*}
\tabcolsep=4pt
\caption{Radio galaxy sample with \hi\ 21-cm absorption from literature. \label{tab:litgals}}
\centering
\begin{tabular}{lcccccccccllc}
\hline\hline             
Source name  & $z_{em}$(opt)  & $z_{abs}$ & V$_{null, B}^{\dag}$ & V$_{null, R}^{\dag}$ & S$_{1.4\rm GHz}$ & $\log \rm L_{1.4 GHz}$ &  S$_{21-cm}$ & $\int \tau\ dv$   & 21-cm Ref. & LAS & LLS & Ref.\\
  &  & & (\kms\,) & (\kms\,) & (mJy) & (\,W\,Hz$^{-1}$) &  (mJy) & (\kms\,) &  & (arcsec) & (kpc) & \\ 
(1) & (2) & (3) & (4) & (5) & (6) & (7) & (8) & (9) & (10) & (11) & (12) & (13) \\
\hline
J0025$-$2602 & 0.322 & 0.322  & $-$349  & 115     & 8.753 & 27.5 &  8.17  & 1.33  & 1 & 0.65  & 3.02  & 3 \\
J0141$+$1353 & 0.621 & 0.620  & $-$225  & $-$141  & 2.74  & 27.7 &  3.78  & 0.64  & 1 & 0.99  & 6.71  & 3 \\
J0146$-$0157 & 0.959 & 0.959 & $-$700 & 1115 & 882.5 & 27.65 & 1803.8 & 11.46 & 6 & 0.04 & 0.326 &  9\\
J0431$+$2037 & 0.219 & 0.219  & $-$54   & 92      & 3.756 & 26.7 &  4.56  & 2.00  & 1 & 0.29  & 1.02  & 3 \\
J0834$+$5534 & 0.242 & 0.240  & $-$672  & $-$152  & 8.283 & 27.1 &  7.22  & 0.61  & 1 & 11.0  & 41.4  & 3 \\
J0901$+$2901 & 0.194 & 0.194  & $-$161  & 131     & 2.004 & 26.3 &  2.006 & 0.06  & 1 & 5.7   & 18.2  & 3 \\
J1013$+$2448 & 0.950 & 0.947 & $-$832 & $-$287 & 541.7 & 27.33 & 891.9 & 1.15 & 6 & 0.049 & 0.396 & 8 \\
J1048$+$3538 & 0.846 & 0.847 & $-$411 & 132 & 309.3 & 27.03 & 552.9 & 4.77 & 6 & 0.03 & 0.236 & 9\\
J1206$+$6413 & 0.371 & 0.372  & 130     & 407     & 3.720 & 27.3 &  3.34  & 1.13  & 1 & 1.3   & 6.6   & 3 \\
J1247$+$4900 & 0.207 & 0.207  & $-$757  & 61      & 1.140 & 26.1 &  1.226 & 0.46  & 2 & 2.64  & 5.4   & 5 \\
J1326$+$3154 & 0.370 & 0.368  & $-$760  & $-$218  & 4.862 & 27.3 &  5.405 & 0.41  & 1 & 0.056 & 0.285 & 3 \\
J1357$+$4354 & 0.646 & 0.645  & $-$585  & 266     & 0.704 & 27.0 &  0.50  & 19.53 & 1 & 0.017 & 0.117 & 3 \\
J1400$+$6210 & 0.431 & 0.430  & $-$470  & $-$65   & 4.308 & 27.5 &  5.47  & 1.10  & 1 & 0.065 & 0.362 & 3 \\
J1648$+$2224 & 0.823 & 0.824 & $-$68 & 262 & 253.0 & 26.66 & 252.8 & 16.06 & 6 & 0.0056 & 0.044 & 7\\
J1821$+$3942 & 0.798 & 0.793  & $-$923  & $-$672  & 3.507 & 28.0 &  5.17  & 0.96  & 1 & 0.47  & 3.53  & 3 \\
J1944$+$5448 & 0.263 & 0.257  & $-$1782 & $-$1075 & 1.754 & 26.6 &  1.95  & 2.88  & 1 & 0.040 & 0.161 & 3 \\
J2052$+$3635 & 0.355 & 0.355  & $-$168  & $-$72   & 5.144 & 27.2 &  4.5   & 4.24  & 1 & 0.060 & 0.297 & 3 \\
J2316$+$0405 & 0.220 & 0.219  & $-$405  & $-$803  & 4.676 & 26.8 &  4.97  & 0.43  & 1 & 1.86  & 6.55  & 4 \\
J2355$+$4950 & 0.238 & 0.238  & $-$111  & 150     & 2.306 & 26.6 &  2.25  & 1.66  & 1 & 0.069 & 0.256 & 3 \\
\hline
\hline
\end{tabular}
\tablebib{
(1) \cite{Vermeulen03}; (2) \cite{Maccagni17}; (3) \cite{Gupta06}; (4) \cite{Grasha19}; (5) \cite{Sanghera95}; (6) \cite{Aditya2019}; (7) \cite{Helmboldt07}; (8)~\cite{Deller2014}; (9) \href{http://astrogeo.org/calib/search.html}{C-band~map~from~VLBI~Calibrator~search}
}
\tablefoot{
\small{
    Column\,1: Source name (J2000); Column\,2: Optical redshift; Columns\,3: Peak \hi\ 21-cm absorption redshift; Column\,4 - 5: Velocity to the bluer and redder side of the peak, respectively, where optical depth approaches zero; Column\,6: Flux density at 1.4 GHz; Column\,7: Radio luminosity at 1.4\,GHz; Column\,8: Flux density at the redshifted 21-cm line frequency; Column\,9: Total velocity integrated optical depth; Column\,10: \hi\ 21-cm absorption reference; Column\,11: Largest Angular Size in arcsec, and Column\,12: Largest Linear Size in kpc; Column\,13: Reference for LAS and LLS.\\}
    \footnotesize{$^{\dag}$ measured with respect to $z_{em}$(opt).\\}
}
\end{sidewaystable*}

\end{appendix}

\end{document}